\documentclass[a4paper,twoside,twocolumn]{article}
\pdfoutput=1

\newif\iftr 
\trtrue     
\newif\ifccs   
\ccstrue      

\usepackage{ifpdf}  
\ifpdf
    \usepackage[pdftex]{graphicx}
    \usepackage[colorlinks,linkcolor=blue,citecolor=blue,urlcolor=blue]{hyperref}
    \DeclareGraphicsExtensions{.pdf}
\else
    \usepackage{graphicx}
    \usepackage[hypertex]{hyperref}  
\fi

\usepackage{todonotes} 

\usepackage{
url       %
,fancyhdr 
,lastpage 
,enumerate 
,amsmath  
,amsfonts 
,amssymb  
, enumitem 
, array 
}
\usepackage[hang]{footmisc} 
\usepackage[noindent, arraystretch, fullpage]{setlengths} 
\usepackage{
own         
}

\graphicspath{{images/}}

\newcommand*{\metaauthori}{Bob Briscoe}

\newcommand*{\metashorttitle}{Managing a Queue to a Soft Delay Target}
\newcommand*{\metatitle}{{\large{Technical Report}}\\Managing a Queue to a Soft Delay Target}
\newcommand*{\metano}{TR-BB-2017-003}
\newcommand*{\metakeywords}{Data Communication, Networks, Internet, Control, Congestion Control, Quality of Service, Performance, Latency, Responsiveness, Dynamics, Algorithm, Standards, Active Queue Management, AQM, Signalling, Sojourn time, Queue delay, Service time, wait time, Expectation, Estimation, Explicit Congestion Notificiation, ECN}

\newcommand*{\metamaili}{\href{mailto:research@bobbriscoe.net}{research@bobbriscoe.net}}

\newcommand*{\metaaddress}{}

\newcommand*{\metaversion}{02}
\newcommand*{\metadate}{15 Apr 2019}

\hypersetup{                       
     pdfauthor = {\metaauthori%
     },
     pdftitle = {\metashorttitle},
     pdfsubject = {},
     pdfkeywords = {\metakeywords}
}%

\title{\metatitle}%
\author{\metaauthori%
\thanks{\metamaili, %
\metaaddress}%
}
\date{\metadate}%

\fancypagestyle{first}{%
\fancyhead[LO,RE]{}%
\fancyhead[LE,RO]{}%
\fancyhead[C]{}%
}%

\begin{document}
\bibliographystyle{alpha}%


\maketitle%
\thispagestyle{first}

\begin{abstract}
{\small\noindent%
This memo proposes to transplant the core idea of Curvy RED, the softened delay target, into AQMs that are better designed to deal with dynamics, such as PIE or PI2, but that suffer from the weakness of a fixed delay target.

}      
\end{abstract}
\ifccs{}%
%
%
\subsection*{CCS Concepts}
\textbf{\textbullet Networks} \(\to\) \textbf{Cross-layer protocols; Network algorithms; Network dynamics;}
\subsubsection*{Keywords}
\metakeywords
\fi{}%


\section{Introduction}\label{softtargettr_intr}

All the well-known modern AQMs aim for a constant target queueing delay, e.g. CoDel~\cite{Nichols12:CoDel} including fq-CoDel, PIE~\cite{Pan13:PIE}, PI\(^2\)~\cite{DeSchepper16a:PI2}, DualPI\(^2\)~\cite{Briscoe15e:DualQ-Coupled-AQM_ID}, and a recent variant of DCTCP's AQM~\cite{Bai16:MQ-ECN, Bai16:ECN_GPS}. In the technical report ``Insights from Curvy RED''~\cite{Briscoe15b:CRED_Insights} it was proved that the level of loss needed to induce TCP-based load to keep to a fixed delay target has to rise to unacceptably high levels during periods of increased load.

The report makes the point that the time taken to repair losses is itself a source of delay, particularly for short flows. Therefore, it is perverse to hold down queuing delay at the expense of very high loss levels.

\cite{Briscoe15b:CRED_Insights} proposes an adaptation of the RED algorithm~\cite{Floyd93:RED} called Curvy RED that uses a convex function of queuing delay as a target. This softens (increases) the delay target as load intensifies. 

However RED, and by extension Curvy RED, provides no control over queue dynamics, whereas control theoretic AQMs do. This means that during dynamic load excursions, RED and Curvy RED have little control over how much delay overshoots (or undershoots) while trying to bring it back to the target. This allows delay to vary uncontrollably above the target. This is a significant problem because many applications are sensitive to maximum, not average, delay.

This memo proposes to transplant the core idea of Curvy RED, the softened delay target, into AQMs that are better designed to deal with dynamics, such as PIE or PI2, but that suffer from the weakness of a fixed delay target.

It should be emphasized that this combination is only useful when loss is used as the signalling mechanism. By extension that means this combination would also be used with classic ECN~\cite{IETF_RFC3168:ECN_IP_TCP}, which requires any ECN behaviour to be equivalent to loss behaviour. However, this combination would be unnecessary for use with L4S ECN~\cite{Briscoe15f:ecn-l4s-id_ID},  which is not constrained to be equivalent to loss.


\section{Curvy PI\(^2\)}\label{softtargettr_curvy-pi}

A Proportional Integral (PI) controller alters the congestion signalling probability dependent on both the distance from the target delay (the error) and the rate of change of the queuing delay. Therefore, it is able to rapidly control load excursions before they cause too much variation in load. Specifically, it uses an control equation of the form:
\begin{equation*}
	p^\prime(t) = p^\prime(t-T) + \alpha\big(q(t)-Q_0\big) + \beta\big(q(t)-q(t-T)\big),\label{eqn:pi}
\end{equation*}
where \(p^\prime(t)\) is the drop probability at time \(t\), \(q(t)\) is the queuing delay at time \(t\),  \(\alpha\) and \(\beta\) are the gain constants, \(T\) is the sampling period and \(Q_0\) is the (constant) target delay.

It would be straightforward to make the target delay a function \(Q()\) of the current drop probability \(p(t-T)\) rather than a constant, for instance, picking a reasonable formula fairly arbitrarily, one might use:
\begin{equation}
	Q(p^\prime) = Q_0 + Q_1 p^\prime(t-T),\label{eqn:softtarget-base}
\end{equation}
where \(Q_0\) and \(Q_1\) are the min and max values of the soft delay target (at \(p^\prime = 0\) and \(p^\prime = 1\)).

The PI\(^2\) controller squares the resulting value of \(p^\prime(t)\) to determine the drop probability \(p = (p^\prime)^2\) (see \cite{DeSchepper16a:PI2} for why). Therefore \autoref{eqn:softtarget} is equivalent to:
\begin{equation}
	Q(p) = Q_0 + Q_1 \sqrt{p(t-T)}.\label{eqn:softtarget}
\end{equation}

\begin{figure}
	\centering
	\includegraphics[width=\columnwidth]{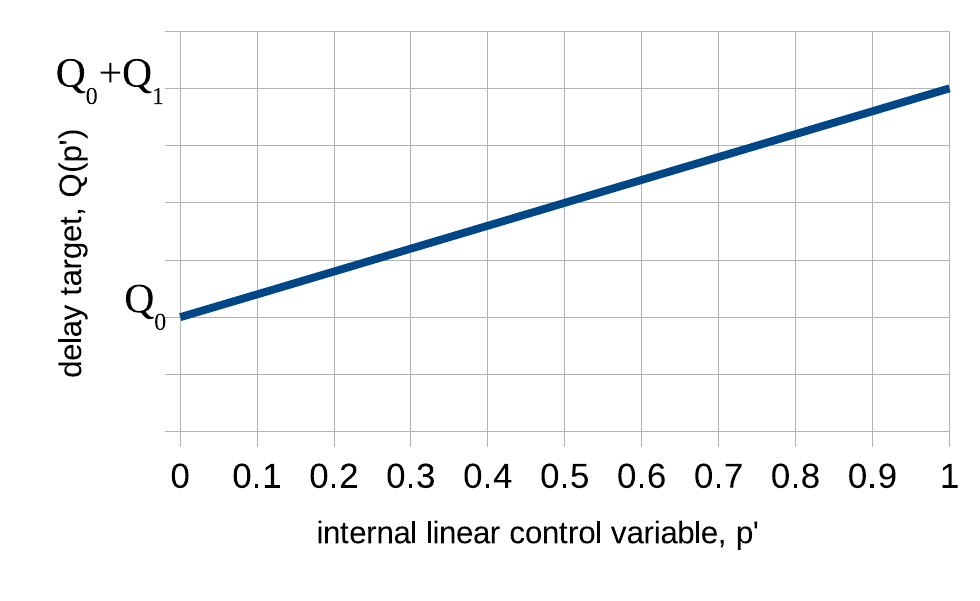}\\
	\includegraphics[width=\columnwidth]{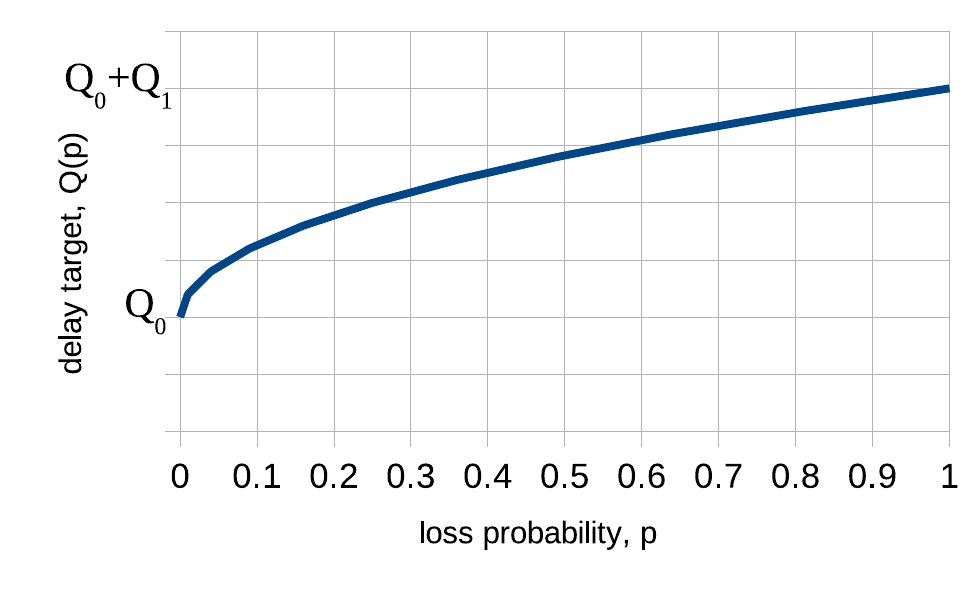}	
	\caption{Soft delay target as a function of \(p^\prime\) \& \(p\)}\label{fig:softtarget}
\end{figure}

Equations \ref{eqn:softtarget-base} and \ref{eqn:softtarget} are illustrated in \autoref{fig:softtarget}

In practice, rather than using the arbitrary formula in \autoref{eqn:softtarget-base}, it will be possible to determine the optimum compromise between queueing delay and loss from human factors experiments that record the mean opinion score of a 2-D matrix of these two impairments for the popular application that is most sensitive to both, e.g. voice, or perhaps virtual reality (although MOS data is more readily available for voice). Then it should be possible to fit an approximate curve to the contour of optimum pairs that will be amenable to implementation as the soft delay target function.

It might seem of concern that the loss probability depends on a delay target function, which in turn depends on the loss probability, which seems like a circular dependency. However, the target function depends on the loss probability that was output in the previous sampling period (and the second dependency will be much weaker than the first anyway).

It might seem contrary to the goal of a PI controller to allow high load to increase delay. However, there is nothing sacred about the constant delay goal that was first proposed by Hollot \emph{et al} in 2001~\cite{Hollot01b:Ctrl_Theoretic_RED}, before designing a solution in the same year~\cite{Hollot01a:PI_AQM}. A controller can aim for any target that meets human needs, it does not have to be a constant.

\section{Variants}\label{softtargettr_variants}

By extension, a delay target that itself depends on the level of loss could be used in other, non-control-theoretic AQMs such as CoDel. As before, the intent would be to soften the delay target under high load, so as not to drive loss to extreme levels in pursuit of low queuing delay, given repairing loss itself introduces delay. 

In the case of Codel, the variable called \texttt{target} would need to depend on the variable \texttt{drop\_next\_}, which determines the interval between drops.


\addcontentsline{toc}{section}{References}

{\footnotesize%
\bibliography{aqm-details}}


\onecolumn%
\addcontentsline{toc}{part}{Document history}
\section*{Document history}

\begin{tabular}{|c|c|c|p{3.5in}|}
 \hline
Version &Date &Author &Details of change \\
\hline\hline
01                      &07 Sep 2017   &Bob Briscoe &First complete version.\\\hline%
\metaversion &\metadate     &Bob Briscoe &Added abstract and keywords.\\\hline%
\hline%
\end{tabular}

\end{document}